# In situ evidence for the structure of the magnetic null in a 3D reconnection event in the Earth's magnetotail

C.J. Xiao[1], X. G. Wang[2], Z.Y. Pu[3*], H. Zhao[1], J.X. Wang[1], Z. W. Ma[4], S.Y. Fu[3], M. G. Kivelson[5], Z.X. Liu[6], Q. G. Zong[7], K.H. Glassmeier[8], A.Balogh[9], A. Korth[10], H. Reme[11], C. P. Escoubet[12]

1. National Astronomical Observatories, Chinese Academy of Science, Beijing 100012, China,

2. State Key Lab of Materials Modification by Laser, Ion, and Electron Beams, Dalian University of Technology, Dalian 116024, China,

3. School of Earth and Space Sciences, Peking University, Beijing 100871, China,

4. Institute of Plasma Physics, Chinese Academy of Science, Hefei 230031, China,

5. Institute of Geophysics and Planetary Physics, University of California, Los Angeles, California, USA,

6. CSSAR, Chinese Academy of Science, Beijing 100080, China,

7. Centre for Space Physics, Boston University, Boston, Massachusetts 02215, USA,

8. IGEP, Technische Universität Braunschweig, Braunschweig, Germany,

9. Department of Physics, Imperial College of Science, Technology and Medicine, London, UK,

10. MPI for Solar System Research, Katlenburg-Lindau, Germany,

11. Centre d'Etude Spatiale des Rayonnements, BP 4346, 31028 Toulouse Cedex 4, France ,

12. ESA/ESTEC, Postbus 299, 2200 AG Noordwijk, The Netherlands

* Corresponding author

**Abstract:** Magnetic reconnection is one of the most important processes in astrophysical, space and laboratory plasmas. Identifying the structure around the point at which the magnetic field lines break and subsequently reform, known as the magnetic null point, is crucial to improving our understanding reconnection. But owing to the inherently three-dimensional nature of this process, magnetic nulls are only detectable through measurements obtained simultaneously from at least four points in space. Using data collected by the four spacecraft of the Cluster constellation as they traversed a diffusion region in the Earth's magnetotail on 15 September, 2001, we report here the first in situ evidence for the structure of an isolated magnetic null. The results indicate that it has a positive-spiral structure whose spatial extent is of the same order as the local ion inertial length scale, suggesting that the Hall effect could play an important role in 3D reconnection dynamics.

Magnetic reconnection is considered to be of crucial importance not only to space science but also to magnetically confined fusion studies and many basic processes in astrophysics and cosmology. The very concept of reconnection was at first suggested at a topologically distinct location called a null point, where the magnetic field is annihilated and magnetic energy is dissipated [1, 2]. In magnetospheric plasmas, reconnection configurations with nulls in two-dimensional (2D) geometry, i.e., X-points, for southward or northward interplanetary magnetic field (IMF), have been proposed and studied for many years [3-12]. For an arbitrary IMF orientation however, reconnection is predicted to occur along a field line linking a pair of nulls in three-dimensional (3D) geometry, called the separator, and separatrix surfaces, called Σ-surfaces or preferably fans [13], generated by the nulls [13-15]. Observations in solar and other astrophysical plasmas also indicate that reconnection is essentially three dimensional [16-19]. On the other hand, nulls and their 3D properties are as well of crucial importance in topology, nonlinear dynamics and the interaction of complex fields [20, 21]. It is therefore necessary to go beyond the well developed 2D literature to 3D reconnection studies.

Steady state and kinematic fluid approaches have been developed in 3D reconnection studies, and a few possible topologies of reconnection with or without nulls have been discussed [15-18, 22-24]. Nevertheless, detailed study of 3D reconnection dynamics with nulls started only recently [25,26]. Furthermore, though null points have been reported in several simulation results [18, 27-29], it is a hard task to find them either in satellite observations or in laboratory experiments. Certain pioneering attempts to understand the physics of 3D nulls have been made in solar plasma studies [13, 29-35]. Since the solar magnetic field measurement is currently limited to the photosphere surface, indirect methods have been developed to search for magnetic nulls three dimensionally, such as from the observed chromospheric and coronal intensity structures [31] or extrapolating the 3D magnetic configurations above the photosphere by solving the boundary problem under either a current-free, or force-free approximation [32-34]. Magnetic field measurement technique in laboratory plasmas is still under development [36]. Thus, the Earth's magnetosphere presents a most promising site for direct observation and characterization of 3D magnetic nulls.

Though providing some evidences for reconnection features [8-12], single-satellite data are not useful for in-situ null detection, since geometrically the topological measure of a null point in space is zero and at least 4-point measurements are needed to characterize 3D structures. With four identical spacecraft in elliptical orbits of 4×19.6 Earth radii ($R_E$) and spacing in a range of hundreds to thousands of kilometers, Cluster [37] is explicitly designed for studying small-scale 3D structures and multi-scale dynamics in key regions of the magnetosphere. The mission thus provides the first opportunity to detect a magnetic null point in the reconnection region of magnetospheric plasmas. We report here for the first time the in situ evidence for a 3D magnetic null by the Cluster constellation. Important topological and physical properties of the null are also revealed by the high resolution data of the magnetic field.

The event lasted from 05:00 to 05:15 UT on 15 September 2001, as Cluster traversed the central current sheet of the magnetotail at ~ (-18.7, 3.5, -2.9) $R_E$ (GSM coordinates) [38]. While C3 remained south of the current sheet, C1, C2 and C4 meandered through the diffusion region several times. The magnetic field (***B***) and plasma velocity (***V***) data of the FGM and CIS onboard Cluster [39, 40] (no ***V*** measurement for C2) during the interval 05:03-05:04 UT are plotted in Figure 1. C1 and C4 first detected earthward plasma jets ($V_x>0$, $B_z>0$) followed by tailward jets ($V_x<0$, $B_z<0$). A distinct quadrupolar Hall magnetic field component $B_y$ was clearly present. All these characteristics are in a good agreement with a crossing of the typical diffusion region in collisionless reconnection [7] (see the schematic illustration in Figure 2a). Of particular interest is the position of four satellites around the X-point of 2D projection of the reconnection configuration at 05:03:36 UT, shown in Figure 2a. Based on reconnection theories, if there exists a null point in 3D geometry, it is most likely to be located within the tetrahedron formed by the four spacecraft (Figure 2b).

The fact that the magnetic null ($\mathbf{B}=0$) is a mathematically isolated neutral point makes it hard to identify the null from magnetic field data. First, right at the null, the relative measurement error is infinite. Secondly, in principle a single point measurement cannot distinguish an isolated null from a magnetic neutral line or a neutral sheet, since the null is always associated with the topological property of its neighborhood. In general a null point is structurally or topologically stable (in the sense that it is preserved when the field is perturbed), whereas a neutral sheet or line is not [16]. To distinguish the null, there are a few methods in the literature, such as the Poincaré index and linear interpolation methods, first introduced to plasma physics studies by Greene [22, 41] then developed by others [34, 42], to characterize a null from data around it. We then first identify the null point using the topological degree method to calculate the Poincaré index, the winding number of a vector field in global differential topology [41]. The topological degree of a magnetic field can be defined as follows. The magnetic field $\mathbf{B}(x,y,z)=(B_x,B_y,B_z)$ can make a mapping from the configuration space $(x,y,z)$ to an "M" space $(B_x,B_y,B_z)$. Then clearly an isolated magnetic null will be mapped as the origin of "M" space, and any closed surface in the configuration space can be mapped as a "balloon", e.g., a closed manifold in the "M" space. Thus if a closed surface in the configuration space surrounds an isolated singular point, the solid angle subtended by the corresponding "balloon" in the "M" space should be 4π. If there is no such a singular point, it is zero [34, 41]. The topological degree, therefore the Poincaré index, then can be the solid angle (with the direction of the surface) normalized by 4π. In certain previous works [34, 41], the cubic surface was used to as the closed surface in the configuration space to calculate the Poincaré index. For Cluster measurement with only four satellites, we have to use a tetrahedron, composed of four triangles, with the four spacecraft being its vertices, as the closed surface (Figure 2b), and calculate the Poincaré index with simultaneous 4s-resolution magnetic field data. The solid angle subtended by the "M" space mapping of any one of the four triangles can be determined by the magnetic field vectors on its three vertices. Summing up the four solid angles corresponding to the four triangles of the tetrahedron and divided it by 4π, we obtain the Poincaré index. Figure 2c shows that the calculated Poincaré index at 05:03:36 UT is 0.9793, very close to unity. Using intercalibrated simultaneous high resolution (0.04s) magnetic field data (shown Figures 3a-3d) from 05:03 to 05:04 UT, we find that the Poincaré index becomes unity in a brief interval around 05:03:36 (see Figure 3e). On the other hand, outside this interval the Poincaré index is always found to be less than $10^{-6}$. These results demonstrate the evidence of an isolated null point existing within the tetrahedron. As a reference, a linear interpolation method is then also applied [41]. At 05:03:36 UT, a point with $\mathbf{B} = 0$ is indeed found inside the tetrahedron at (85.3, -81.5, 901.5) km relative to C3, shown in Figure 2b, reconfirming the evidence for the existence of the null point inside the tetrahedron.

Note that the topological degree method requires the closed surface to approach the null and the magnetic field inside it to be continuous, without significant variations. Furthermore, the more data points are used, the better result it makes. Therefore the finite distance to the null and noteworthy nonlinear changes of the field, as well as the data from only four points, the least number needed, would cast a certain degree of uncertainty on the calculation. Nonetheless, Cluster measurement is the best mean for three-dimensionality up-to-date. On the other hand, the numerical calculation should also bring an error bar to the result so that the index would not be exact integers. In this event however the error is about 2%.

The high resolution magnetic field data are also used to further analyze the 3D structure in the vicinity of the null.

Shown in Figure 3a ($B_y$), 3b ($B_z$) and 3d ($B_z$), a bipolar magnetic structure was detected during the interval 05:03:33-05:03:40 UT. The bipolar structure may indicate a satellite crossing of either a reconnection location or a flux tube. Since the crossing occurs immediately after the plasma flow reversal from earthward to tailward and with a positive $B_x$, i.e., the Cluster spacecraft moved from Quadrant (II) to (I), it is clear that the flux tube is only located in Quadrant (I) of the diffusion region marked in Figure 2a where $B_x>0$ and $V_x<0$. Therefore it is unlikely that the bipolar structure shows a reconnection site traversing, but a flux tube.

The bipolar structure cannot be explained as a local flux rope formed in 2D reconnection, either. As shown in Figure 4a, since the satellites are traveling tailward relatively to the structure, in the magnetotail, if it were a flux rope crossing in 2D reconnection frame, then should the $B_z$ component of the bipolar structure have changed from negative to positive, while the direction of the $B_y$ component should not have changed. The observed bipolar component shown in Figures 3a , 3b , and 3d however always changes from positive to negative. This is no doubt a great challenge to conventional 2D reconnection models. The other challenge is the axis orientation of the flux tube, which is typically in the y-direction for 2D reconnection in the magnetotail. Due to the relative satellite position in the x-direction as shown in Figure 2a, such a 2D bipolar structure with a y-direction axis would have been first detected by C4, then C2 and C1, as shown in Figure 4a. However, the tube in fact is detected first by C2, and then by C4 and C1 three seconds later (see Figure 3a, 3b and 3d). Applying the minimum variance analysis (MVA) method [43] to C1, C2 and C4 data, we are able to show in Figure 4c that the axis orientation of the flux tube lies basically in the x-z plane, but not in the y-direction ($\phi_{C1}=206.1^0$, $\theta_{C1}=40.5^0$; $\phi_{C2}=199.7^0$, $\theta_{C2}=71.5^0$; $\phi_{C4}=185.8^0$, $\theta_{C4}=90.2^0$ in GSM coordinates). Meanwhile, during the flux tube crossing of the Cluster satellites, the dominant plasma flow is almost anti-parallel to the y-axis (e.g., $V_{4x}=$ -119.5 $km/s$, $V_{4y}=$ -605.2 $km/s$, $V_{4z}=$ 72.6 $km/s$ as shown in Figure (1b) and (1g)), the perpendicular (to the flux tube axis) component of $V_4$ in the y-direction is $V_{4\perp,y}$ =447.8km/s. Therefore the distance of $\Delta y_{24} =| y_4 - y_2 |= 1570.8 km$ could be passed by $V_{4\perp,y}$ in about 3.5 seconds. In fact, C4 sees the structure ~3 seconds later than C2, as shown in Figure 4b. This strongly suggests a y-direction crossing, but not the x-direction crossing. The relative motion of C1 is similar to C4.

Thus, to understand the observed features of the flux tube, we combine the axis orientation analysis with the spiral property of the flux tube characterized by the polarity of the bipolar component variation presented in Figure 4c. We can then infer a 3D structure in the vicinity of the null as shown in the cartoon of Figure 4d. To further understand this structure and its spiral orientation, a topological model can be applied. In the model the topological property of the null can be obtained from the eigenvalues of the matrix $\delta \mathbf{B} \equiv \nabla_j B_i = \partial B_i / \partial x_i$ [13, 15, 20, 28-30]. At least four satellites are necessary to characterize the three dimensional features of the magnetic field because there are only three coefficients for the magnetic field at a point (e.g. at one of the satellites) and nine are needed to compute the matrix $\nabla_j B_i = \partial B_i / \partial x_i$. Also $\nabla \cdot \mathbf{B} = 0$ adds another constraint. Based on simultaneous 4s-resolution magnetic field data of the four-spacecraft at 05:03:36 UT, the matrix $\delta \mathbf{B}$ can be calculated using a linear interpolation approximation [44-46]

$$\delta \mathbf{B} = \begin{bmatrix} -0.0012550, & -0.0110685, & 0.0232983 \\ 0.0090147, & 0.0015492, & 0.0009428 \\ 0.0019898, & -0.0047371, & -0.0002319 \end{bmatrix},$$

with eigenvalues of $\lambda_1 = 0.0043 + 0.0105i$, $\lambda_2 = 0.0043 - 0.0105i$, $\lambda_3 = -0.0086$. Clearly the 3D isolated null in the diffusion region observed at 05:03:36 UT is the positive spiral type [13] with a spine in the x-z plane (determined by eigenvector of $\lambda_3$). Of course there are still uncertainties in the result due to both the measurement error and the linear interpolation approximation. The intercalibration of the magnetometers is good to 0.1 nT [39], and in a field of the order 10-20 nT, the accuracy of FGM measurement is about 1 part in 100 or 200. To include the error brought by the linear approximation on the other hand, we apply the constraint $\nabla \cdot \mathbf{B} = 0$ as a test of the accuracy. In this event it is up to $10^{-5}$ based on $\nabla \cdot \mathbf{B} = 0.0000623$ nT/km. The total relative error in calculating $\delta \mathbf{B}$ can be characterized by the ratio of $|\nabla \cdot \mathbf{B}| / |\nabla \times \mathbf{B}|$ [44-46]. In this event it is $|\nabla \cdot \mathbf{B}| / |\nabla \times \mathbf{B}| \approx 2.8\%$. Clearly, within this accuracy level, the characteristics of the null point are reliable, because they only depend on the signs of the eigenvalues, particularly the real eigenvalue ($\lambda_3$ in this paper). To the Poincaré index calculation above, such an accuracy level is also good enough since the index is either unity with a null, or zero with no null.

Figure 4c shows a sketch of the separatrix surface structure and the spine of the positive spiral type null [13, 15, 22]. The axis orientation of the flux tube calculated above in MVA from C1, C2 and C4 data matches the spine coincidentally. The geometrical structure of the magnetic field lines around the positive spiral type null is schematically illustrated in Figure 4d, where the magnetic field lines spiral around the spine and form a flux tube. Obviously the theoretical structure of the null is in agreement with the bipolar polarity and axis orientation of the flux tube very well. Therefore the bipolar structure can be naturally identified as the spiraling field lines generated by the null along the spine. Also significantly, in the above observation the maximum cross section size of the flux tube, i.e., the width of the localized current density profile, can be estimated as ~1500 km for a bipolar structure crossing interval of ~2.5 seconds with a crossing velocity of ~600 km/s. Then the upper limit of the spine flux structure size is about the order of local ion inertial length $d_i$ ($\approx 509 km$ in this case [38]).

The existence of nulls is of fundamental significance to the 3D reconnection process. It is also of crucial importance in vector field topology and the interaction of complex differential manifolds. We have used full 3D in situ measurements from the four Cluster spacecrafts at high time resolution to reveal the evidence of a magnetic null in the geomagnetotail. The challenge in identifying a 3D null is theoretically the geometrical measure of a spatial null is zero. Nevertheless, we have been able to show the evidence of the null by using the Poincaré index. We have also been able to characterize the properties of the local structure through analysis of the magnetic difference matrix, $\delta\mathbf{B}$, concluding that the magnetic null is a positive spiral type point. Besides the topological features, the analysis of spacecraft in-situ measurements also shows that the characteristic size of the spine structure is about the order of the local ion inertial length, a result previously reported neither in observations nor in theory and simulations. This ion inertial scaling of the null structure shows that, similar to 2D reconnection, 3D reconnection in collisionless plasmas is most probably similar to 2D reconnection in being dominated by the Hall effect featured by the $d_i$-scaling [47, 48]. It should be fundamental to the dynamics of 3D collisionless reconnection around nulls.

With only one null point identified, we can not establish if there are any other nulls or null-null lines existing in the diffusion region nor can we suggest which 3D reconnection model with nulls, namely fan reconnection, spine reconnection or separator reconnection [13, 15-17, 22], is present in this event. Furthermore, lacking of high-resolution plasma data, we must look forward to future work to characterize plasma dynamics in 3D reconnection around the null.

………………………………………………………………

References:

1.  Giovanelli, R. G. A theory of chromospheric flares. *Nature* 158, 81-82 (1946).

2.  Dungey，J. W. *Cosmic Electrodynamics* (Cambridge University Press, Cambridge, 1958).

3.  Dungey, J. W. Interplanetary magnetic field and the auroral zones. *Phys. Rev. Lett.* 6, 47-48 (1961).

4.  Dungey，J. W. in *Geophysics, The Earth's Environment* (ed. Dewitt, C. et al.) 526-531 (Gordon and Breach, New York, 1963).

5.  Parker, E. N. Sweet's mechanism for merging magnetic fields in conducting fluid. *J. Geophys. Res.* 62, 509-520 (1957).

6.  Petschek, H. E. in *The Physics of solar Flare* (ed. Hess, W. N.) 425-437 (NASA SP-50, Washington DC, NASA, 1964).

7.  Birn, J. et al. Geospace environmental modeling (GEM) magnetic reconnection challenge. *J. Geophys. Res.* 106, 3715-3719 (2001).

8.  Phan, T. D. et al. Extended magnetic reconnection at the earth's magnetopause from detection of

bi-directional jets. *Nature* 404, 848-850 (2000).

9. Deng, X. H. & Matsumoto, H. Rapid magnetic reconnection in the Earth's magnetosphere generated by whistler waves. *Nature* 410, 557-559 (2001).

10. Øieroset, M. et al. In situ detection of collisionless reconnection in the earth's magnetotail. *Nature* 412, 414-417 (2001).

11. Mozer, F. S., Bale, S. D. & Phan, T. D. Evidence of diffusion regions at a subsolar magnetopause Crossing. *Phys. Rev. Lett*. 89, 015002 (2002).

12. Frey, H. U., Phan, T. D., Fuselier, S. A. & Mende, S. B. Continuous magnetic reconnection at Earth's magnetopause. *Nature* 426, 533-536 (2003).

13. Priest, E. R. & Titov, V. S. Magnetic reconnection at three-dimensional null points. *Phil. Trans. R. Soc. Lond. A* 354, 2951–2992 (1996).

14. Cowley, S. W. H. A qualitatively study of the reconnection between the Earth's magnetic field and an interplanetary field of arbitrary orientation. *Radio Science* 8, 903-913 (1973).

15. Lau, Y.-T., & Finn, J. M. Three-dimensional kinematic reconnection in the presence of field nulls and closed field lines. *Astrophys. J.* 350, 672 (1990).

16. Priest, E. R. & Forbes, T. G. *Magnetic Reconnection: MHD theory and applications*. (Cambridge Univ. Press, New York, 2000).

17. Priest, E. R., Hornig, G. & Pontin, D. I. On the nature of three-dimensional magnetic reconnection. *J. Geophysics. Res.* 108, doi:10.1029/2002JA009812 (2003).

18. Buchner, J. Three-dimensional magnetic reconnection in astrophysical plasma – kinetic approach. *Astrophys. and Space Sci.* 264, 25-42 (1999).

19. Lui, A. T. Y. Critical issues on magnetic reconnection in space plasmas. *Space Sci. Rev.* 116, 497-521 (2005).

20. Arnold, V. I. *Ordinary differential equation.* (Springer-Verlag, Berlin, New York, 1992).

21. Wang, J. Vector magnetic field and magnetic activity on the Sun. *Fund. Cosmic Physics* 20, 251-382, (1999).

22. Greene, J. M. Geometrical properties of 3D reconnecting magnetic fields with nulls. *J. Geophys. Res.* 93, 8583-8590 (1988).

23. Schindler, K., Hesse, M. & Birn, J. General reconnection, parallel electric fields and helicity. *J. Geophys. Res.* 93, 5547-5557 (1988).

24. Wang, X. G. & Bhattacharjee, A. A three-dimensional reconnection model of the magnetosphere: Geometry and kenematics. *J. Geophys. Res.* 101, 2641-2653 (1996).

25. Greene, J. M. & Miller, R. L. in *Proceedings of the International Symposium in Honor of Bruno Coppi* (Report GAA21961, General Atomics, San Diego. 1995).

26. Hu, S., Bhattacharjee, A., Dorelli, J. & Greene, J. M. The spherical tearing mode**.** *Geophys. Res. Lett.* 31, CiteID L19806 (2004).

27. Galsgaard, K. & Nordlund, A. Heating and activity of the solar corona: 3. Dynamics of a low-beta plasma with three-dimensional null points. *J. Geophys. Res.* 102, 231-248 (1997).

28. Lau, Y.-T. & Finn, J. M. Magnetic reconnection and the topology of interacting twisted flux tubes. *Phys. Plasmas* 3, 3983– 3997 (1996).

29. Pontin, D. I., Hornig, G. & Priest, E. R. in *Proceedings of the SOHO 15 Workshop - Coronal Heating* (ed. Walsh, R.W., et al.) 507-523 (ESA Publications Division, 2200 AG Noordwijk, The Netherlands, 2004).

30. Parnell, C. E., Smith, J. M., Neukirch, T., & Priest, E. R., The structure of three-dimensional magnetic neutral points. *Phys. Plasmas* 3, 759–770 (1996).

31. Filippov, B. Observation of a 3D magnetic null point in the solar corona. *Sol. Phys.* 185, 297-309 (1999).

32. Aulanier, G., et al. The topology and evolution of the Bastille day flare. *Astrophy. J.* 540, 1126-1146 (2000).

33. Fletcher, L. et al., Evidence for the flare trigger site and three-dimensional reconnection in multiwavelength observations of a solar flare. *Astrophy. J.* 554, 451-463 (2001).

34. Zhao, H., Wang, J., Zhang, J. & Xiao, C. J. A new method of identifying 3D null points in solar vector magnetic fields. *Chinese J. Astronomy & Astrophysics* 5, 443-447 (2005).

35. Garth, C., Tricoche, X. & Scheuermann, G. in *Proceedings of IEEE Visualization '04* (ed. Rushmeier, H. et al.) 329-336 (IEEE Computer Society, Washington, DC, 2004).

36. Ding, W. X. Measurement of internal magnetic field fluctuations in a reversed-field pinch by Faraday rotation. *Phys. Rev. Lett.* 90 : Art. No. 035002 (2003).

37. Escoubet, C. P., Schmidt, R. & Goldstein, M. L. in *The Cluster and Phoenix Missions,* (ed. Escoubet, C. P., et al.) 11-32 (Kluwer Acad. Publishers, Dordrecht, the Netherlands, 1997).

38. Xiao, C. J., et al. Cluster measurements of fast magnetic reconnection in Earth's magnetotail. *Geohys. Rev. Lett.*, submitted (2006).

39. Balogh, A. et al. in *The Cluster and Phoenix Missions*. (ed. Escoubet, C. P., et al.) 65-92 (Kluwer Acad. Publishers, Dordrecht, the Netherlands, 1997).

40. Reme, H. et al. in *The Cluster and Phoenix Missions*. (ed. Escoubet, C. P., et al.) 303-350 (Kluwer Acad. Publishers, Dordrecht, the Netherlands, 1997).

41. Greene, J. M. locating three-dimensional roots by a bisect ion method. *J. Comp. Phys.* 98, 194-198 (1992).

42. Cai, D.S., Li, Y. T., Ichikawai, T., Xiao, C. J. & Nishikawa, K. Visualization and criticality of magnetotail field topology in a three-dimensional particle simulation. *Earth, Planets and Space* 53    1011 (2001).

43. Sonnerup, B. U. O. & Scheible, M. in *Analysis Methods for Multi-Spacecraft Data.* (ed. Paschmann, G. & Daly, P. W.) 185-220 (ESA Publications Division, Noordwijk, 1998).

44. Khurana, K. K. et al. Accurate determination of magnetic field gradients from four-point vector

measurements-II: use of nutral constraints on vector data obtained from four spinning spacecraft. *IEEE Trans. On Magnetics.* 32, 5193-5205 (1996).

45. Chanteur, G. In *Analysis Methods for Multi-Spacecraft Data* (ed. Paschmann, G. & Daly, P. W.) 349-369 (ESA Publications Division, Noordwijk, 1998).

46. Xiao, C. J., et al. Multiple magnetic reconnection events observed by Cluster: current calculating, *Chinese Journal of Geophysics* 47, 635-643 (2004)..

47. Wang, X.G., Bhattacharjee, A. & Ma, Z.W. Collisionless reconnection: effects of Hall current and electron pressure gradient. *J. Geophys. Res.* 105, 27633 (2000).

48. Wang, X.G., Bhattacharjee, A. & Ma, Z.W. Scaling of collisionless forced reconnection. *Phys. Rev. Lett.* 87, 265003 (2001).

**Acknowledgements** This work is supported by the NSFC Programs (Grant No. 40390150, 4050421, 10233050, 10575018, 40536030, 40425004, and 40228006) and China Key Research Project (Grant No. G200000784), as well as China Double Star-Cluster Science Team. The authors would also thank Dr. D. S. Cai, Prof. E. R. Priest, and Dr. G. P. Zhou for helpful discussions and suggestions, as well as Dr. H. Schwarzl for producing the intercalibrated FGM data.

**Competing interests statement** The authors declare that they have no competing financial interests.

**Correspondence** and requests for materials should be addressed to Z.Y. P. (zypu@pku.edu.cn)

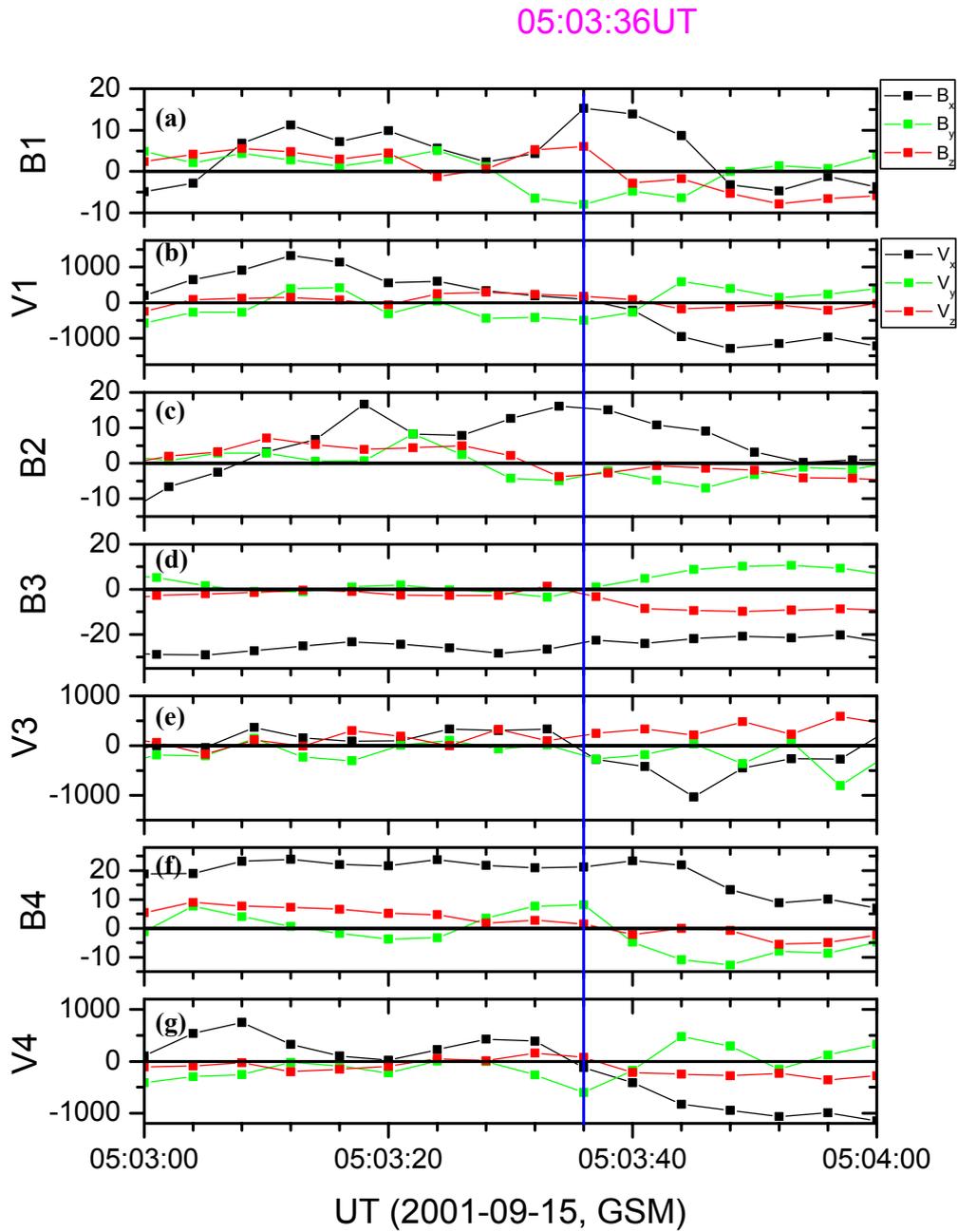

Figure 1: The 4s resolution magnetic field (***B***) and plasma velocity (***V***) data of C1-C4 (no $V_2$ data) during the interval 05:03-05:04 UT on 15 September 2001. Four-spacecraft crossing of a typical diffusion region from earthward ($V_x>0$, $B_z>0$) to tailward ($V_x<0$, $B_z<0$) can be identified. The vertical blue line through all panels at 05:03:36 UT identifies a time of particular interest when the satellite positions surround the reconnecting point ($V_{1x}>0$, $V_{4x}<0$).

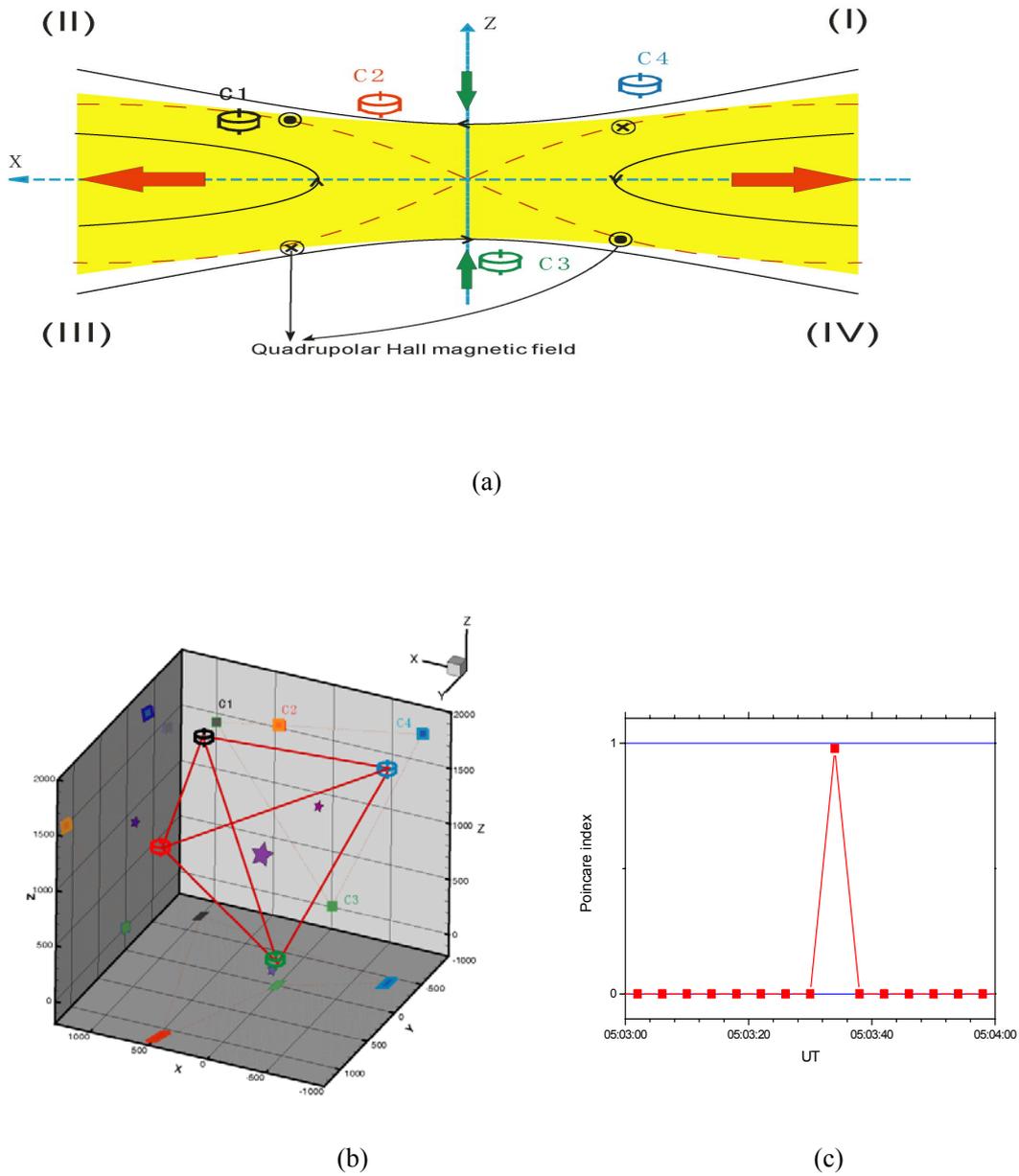

Figure 2: Positions of spacecraft C1, C2 and C4 relative to C3 in GSM coordinates and illustration of the null point surrounding by Cluster satellites. (a) Schematic illustration of a typical geometry of the diffusion region in the (x, z) plane and positions of Cluster spacecraft at 05:03:36 UT, estimated from data in Figure 1. (b) Positions of C1, C2 and C4 at 05:03:36 UT relative to C3 at (-119264.3, 22423.1, -18303.8) km. C1: (1002.5, -750.1, 1419.3) km, C2:

(465.2, 1166.9, 1523.3) km and C4: (-749.8, -403.8, 1741.7) km and the position of the null point (marked by an asterisk) relative to C3 derived from linear interpolation: (85.3, -81.5, 901.5) km. (c) The Poincaré index calculated with simultaneously measured magnetic field data shown in Figure 1.

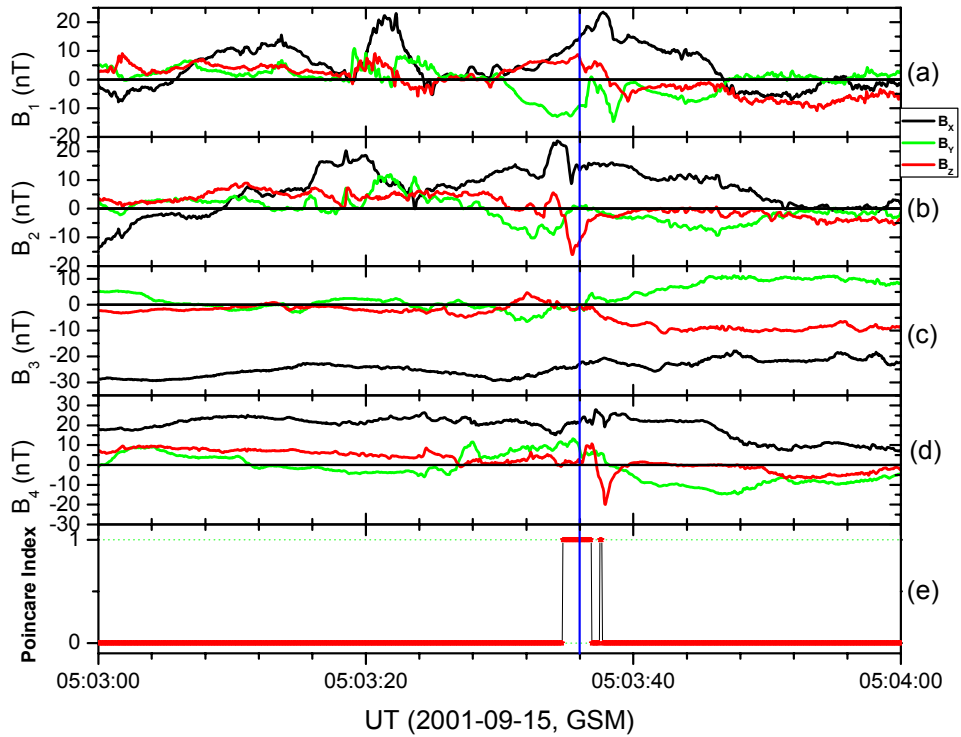

Figure 3: The high resolution (0.04s) magnetic field data of 4 Cluster spacecraft, the calculated Poincaré index during 05:03-05:04 UT on 15 September 2001. (a-d): The magnetic field data of C1-C4. The black, yellow and red indicate the x, y and z component, respectively. (e) The Poincaré index calculated from intercalibrated high resolution (0.04s) data. The vertical blue line shows the time 05:03:36 UT when a null point is found to exist inside the Cluster tetrahedron based on 4s resolution data.

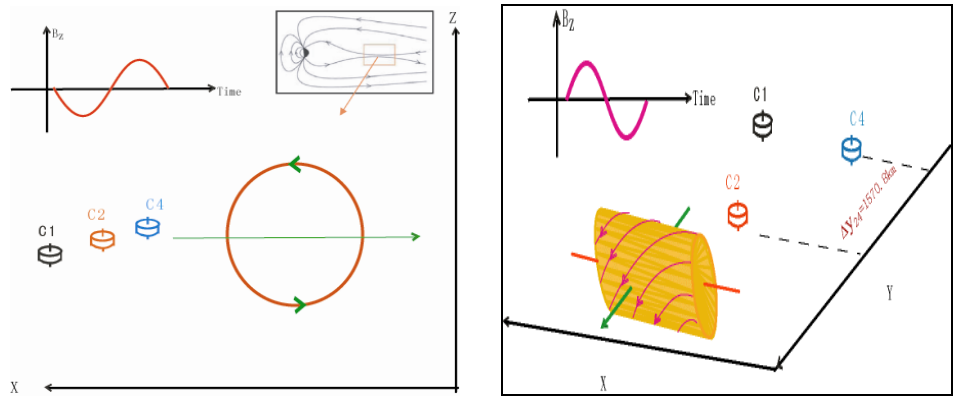

(a) (b)

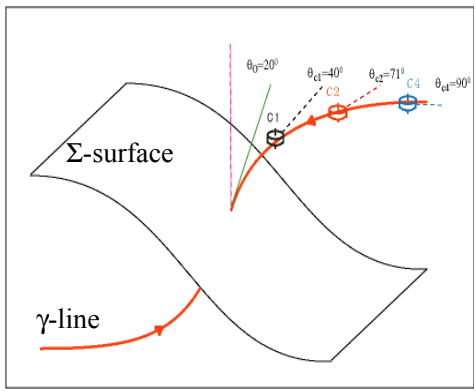
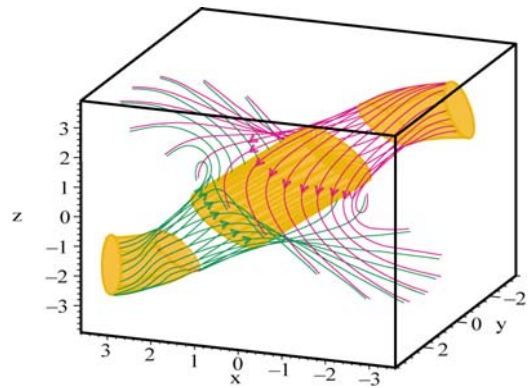

(c) (d)

Figure 4: Cartoons of a positive-spiral-type null and positions of satellites crossing. (a) A schematic illustration of the tailward crossing of a flux tube in a two dimensional model where the polarity of the observed field is shown changing from negative to positive (i.e. a bipolar signature as plotted above left) as the spacecraft motion relative to the structure follows the arrow shown in the lower portion (b) A schematic illustration of the Cluster satellites crossing the flux tube in the

–y-direction (again following the green arrow) where the bipolar signature is shown changing from positive to negative; (c) The axis orientations of the bipolar structure in GSM coordinates, presented in dashed lines and calculated in MVA method from data measured by C1 (black), C2 (red) and C4 (blue) ($\phi_{C1}=206.1^0$, $\theta_{C1}=40.5^0$; $\phi_{C2}=199.7^0$, $\theta_{C2}=71.5^0$; $\phi_{C4}=185.8^0$, $\theta_{C4}=90.2^0$), as well as the orientation of the spine line, shown in the green line and calculated as $\phi_0=169.5^0$, $\theta_0=20.3^0$ by the eigenvector with the real eigenvalue of $\boldsymbol{\delta B}$ matrix at 05:03:36 UT, clearly matching the axis orientations coincidentally. It illustrates a positive-spiral-type null geometry with its fan ($\Sigma$-surface) and spine line ($\gamma$-line). (d) The spiral structure of the magnetic field around the null.